\documentclass[12pt,floatfix,amssymb,amsmath,aps,eqsecnum,nofootinbib]{revtex4}
\usepackage[dvips]{graphicx}
\usepackage{dcolumn}
\usepackage{color}
\begin{document}
\title{Shear viscosity, Bulk viscosity, and Relaxation Times of Causal Dissipative Relativistic Fluid-Dynamics at Finite Temperature
and Chemical Potential}
\author{Xu-Guang Huang$^1$}
\author{Tomoi Koide$^2$}
\affiliation{$^1$ Institut f\"ur Theoretische Physik, J. W. Goethe-Universit\"at,D-60438
Frankfurt am Main, Germany\\
$^2$Instituto de F\'isica, Universidade Federal do Rio de Janeiro, C. P.
68528, 21945-970, Rio de Janeiro, Brasil}

\begin{abstract}
The microscopic formulas for the shear viscosity $\eta$,
the bulk viscosity $\zeta$, and the corresponding relaxation times $\tau_\pi$ and $\tau_\Pi$
of causal dissipative relativistic fluid-dynamics are obtained
at finite temperature and chemical potential by using the projection operator method.
The non-triviality of the finite chemical potential calculation is attributed to the
arbitrariness of the operator definition for the bulk viscous pressure.
We show that, when the operator definition for the bulk viscous pressure $\Pi$ is appropriately
chosen,
the leading-order result of the ratio, $\zeta$ over $\tau_\Pi$, coincides with the
same ratio obtained at vanishing chemical potential.
We further discuss the physical meaning of the time-convolutionless (TCL)
approximation to the memory function, which is adopted to derive the main formulas.
We show that the TCL approximation violates the time reversal symmetry appropriately and leads results  consistent
with the quantum master equation obtained by van Hove.
Furthermore, this approximation can reproduce an exact relation for transport coefficients
obtained by using the f-sum rule derived by Kadanoff and Martin.
Our approach can reproduce also the result in Baier et al.(2008) Ref. \cite{con} by taking into account the next-order correction
to the TCL approximation, although this correction causes several problems.
\end{abstract}

\maketitle

\section{Introduction}

Relativistic hydrodynamics has been widely used to study
various phenomena in many areas, such as heavy-ion collisions,
relativistic astrophysics and cosmology. However,
the formulation of relativistic hydrodynamics is not trivial because
the naive introduction of dissipations violates the relativistic causality,
and leads to inconsistent behaviors in the property of stability \cite{koide_rev,cau_sta}.
Thus the relativistic Navier-Stokes theory contains superluminal modes and
is inadequate as a candidate of relativistic hydrodynamics \cite{herrera}.

Many different approaches have been studied for constructing the consistent relativistic hydrodynamics,
including phenomenological approach \cite{dkkm1,dkkm4, jou, pratt, calzetta, con},
kinetic approach \cite{is, dkr} and others \cite{else}.
The common feature of these approaches is that
the shear stress tensor and the bulk viscous pressure are treated as dynamical variables
in addition to usual hydrodynamic variables (for example, energy density and fluid velocity),
and the evolution of the shear stress tensor (the bulk viscous pressure) follows a
dissipative equation characterized by the shear viscosity (the bulk viscosity)
and the corresponding relaxation time. Different approach predicts different values for the
relaxation times, and also gives different non-linear terms which can
appear in the evolution equations as well.
In the following, we call these theories the causal dissipative relativistic fluid-dynamics (CDRF)
\cite{koide_rev}.

In hydrodynamics, all transport coefficients, such as the shear viscosity and
the bulk viscosity, are inputs and should be determined from a microscopic
theory or experiments.
For the Navier-Stokes theory, the transport coefficients are
usually calculated in two different approaches. One is the kinetic
approach based on the Boltzmann equation, and the other is the
microscopic (field-theoretical) approach through the Green-Kubo-Nakano (GKN) formula.
As known, the applicability of the Boltzmann equation is quite limited and
the GKN formula are more general in application.
However, it should be emphasized that
both approaches give the same results in the classical dilute gas limit \cite{dorfman,footnote}.

Differently from the Navier-Stokes theory, the computing methods for transport coefficients
in CDRF are not well established.
In this work, we will apply the projection operator approach to derive the microscopic formulas for the transport coefficients in CDRF \cite{pro,dhkr,knk,hkkr}, which can be viewed as generalization of the GKN formulas for the Navier-Stokes (Newtonian) fluids.
Although different approaches have been proposed \cite{con,pratt,moore}, as far as we know,
this is the unique approach for which the consistency is confirmed by comparing to the
results obtained from the Boltzmann equation \cite{dhkr}.
Moreover, it is verified that our formula satisfies the exact result obtained from the f-sum rule when applied to the (generalized) diffusion process \cite{koide_dif,hkkr,kadanof}.

The results at the vanishing chemical potential were already reported in Refs.\cite{knk,hkkr}.
The purpose of this paper is to extend this analysis to the case of finite chemical potential.

The non-triviality of the finite chemical potential calculation is attributed to the
arbitrariness of the operator definition of the bulk viscous pressure.
At the vanishing chemical potential, this arbitrariness was removed by comparing with the
result from the Boltzmann equation, but there is no corresponding result at finite chemical potential so far.
We show that, if the bulk viscous pressure is defined appropriately,
the leading-order result of the ratio of the bulk viscosity $\zeta$ to the corresponding relaxation time $\tau_\Pi$ coincides with the same ratio obtained at the vanishing chemical potential.

Another purpose of this paper is to show the physical and mathematical meanings of the
approximation used in the derivation.
In the derivation of the transport coefficients of dissipative equations, we need to introduce  non-trivial
approximations to violate the time reversal symmetry.
As far as we know, there is still no established method for this.
In this paper, we use the so-called time-convolutionless (TCL) approximation.
The meaning of this approximation was already discussed in Ref. \cite{hkkr}.
In this paper, we argue that the quantum master equation obtained by van Hove and the f-sum rule
satisfied in exact quantum time evolution processes are reproduced only when the TCL approximation
to the memory function is applied.
More importantly, in order to keep the consistency between the results of the Boltzmann equation
and quantum field theory we have to use the TCL approximation, as was shown in Ref. \cite{dhkr}.
A naive derivative expansion scheme does not satisfy these conditions.

This paper is organized as follows.
In Sec. \ref{pom}, the projection operator method is briefly summarized.
In Sec. \ref{vio}, the meaning of the TCL approximation is discussed.
By using this approximation, the microscopic expressions for the shear viscosity, the bulk viscosity
and the corresponding relaxation times are derived in Sec. \ref{fcp}.
Sec. \ref{sum} is devoted to the concluding remarks. We use natural units $c=\hbar=k_B=1$.

\section{Projection operator method} \label{pom}

Here we briefly summarize the projection operator method for CDRF \cite{pro,mori,pro2,books}.
The difference between the calculations at vanishing and finite chemical potentials comes from the
different definitions for the projection operators.

Regardless of the presence of the chemical potential,
the time evolution of an arbitrary Heisenberg operator $O$ is determined by the Heisenberg equation of motion,
\begin{equation}
\frac{\partial}{\partial t} O(t) = i [H, O(t)] \equiv iLO(t), \label{hei}
\end{equation}
where $O(t) = e^{iHt}Oe^{-iHt}$ with $H$ the Hamiltonian and $L$ is the Liouville operator.

Let us introduce a set of operators for gross (in our case, the hydrodynamic) variables by ${\bf A}=\{A_i\},\,i=1,\cdots,n$.
In order to implement coarse graining of the Heisenberg equation of motion,
the projection operator is introduced.
Following previous works, \cite{mori,dhkr,knk,hkkr}, we adopt the Mori projection operator,
\begin{equation}
P O = \sum_{i,j=1}^n (O, A^\dagger_j)({\bf A},{\bf A}^\dagger)^{-1}_{ji} {\bf A}_i. \label{moripro}
\end{equation}
The inner product here is given by Kubo's canonical correlation,
\begin{equation}
(X,Y)
= \int^\beta_0 \frac{d\lambda}{\beta} {\rm Tr}[ \rho_{eq} e^{\lambda K}X e^{-\lambda K} Y ],
\end{equation}
where $\rho_{eq} = e^{-\beta K}/{\rm Tr}[e^{-\beta K}]$ with $K = H - \mu N$ and $N$ being a conserved charge. Note that
the statistical expectation is taken not with $H$ but with $K$. It is easy to check that
\begin{equation}
(iLX,Y) = -(X,iLY),
\end{equation}
even for the case of finite chemical potential.

By using this projection operator, we can re-express the Heisenberg equation of motion for the hydrodynamic variables
as follows,
\begin{equation}
\frac{\partial }{\partial t}\mathbf{A}(t)=i\Delta \ \mathbf{A}%
(t)-\int_{0}^{t}d\tau \mathbf{\Xi }(\tau )\mathbf{A}(t-\tau )+\mathbf{\xi }%
(t), \label{morieq}
\end{equation}%
where $\Delta $ and $\Xi $ are $\left( n\times n\right) $
matrices and $\mathbf{\xi }$ is an $n$-vector of operators.
Their elements are given by
\begin{eqnarray}
i\Delta _{ij} &=&\sum_{k}(iL A_{i}, A_{k}^{\dagger })(\mathbf{{%
A}},\mathbf{{A}}^{\dagger })_{kj}^{-1}, \\
\mathbf{\Xi }_{ij}(t) &=&-\theta (t)\sum_{k}(iLQe^{iLQt}iL {A}_{i},
{A}_{k}^{\dagger })(\mathbf{{A}},\mathbf{{A}}^{\dagger })_{kj}^{-1}, \label{xi} \\
{\xi }_{i}(t) &=&Qe^{iLQt}iL {A}_{i},
\end{eqnarray}
where $Q\equiv1-P$.
The first and second terms on the right-hand side represent the collective oscillation
and dissipation after coarse graining of time scale, respectively.
The memory function $\mathbf{\Xi }_{ij}(t)$ can be expressed in terms of the time correlation of the third term, through the fluctuation-dissipation theorem.
If the projection operator is chosen appropriately
so as to collect all the macroscopic degrees of freedom, the third term will oscillate
very fast and can be interpreted as the noise term.
In the following calculation, the noise term is neglected.

\section{Violation of Time Reversal Symmetry, Sum rule and Coarse Graining} \label{vio}

In this section we discuss the approximation which
will be used in the following sections to derive our formulas of transport coefficients.
See also the argument in Sec. V of Ref. \cite{hkkr}.

We have to keep in mind that the evolution equations of viscous fluids
violate the time reversal symmetry, while
the Heisenberg equation of motion (\ref{hei}) is time reversal symmetric.
Thus, to derive the fluid dynamics from an underlying microscopic theory, some non-trivial operations are required. Notice that this is different from the case of the derivation of the fluid dynamics from the Boltzmann equation, because the Boltzmann equation already violates the time reversal symmetry by the assumption of the molecular chaos.
In short, there are two steps to derive fluid dynamics from a microscopic theory:
1) the reduction of the number of the dynamical variables and 2) the violation of time reversal symmetry.
In the Boltzmann equation approach, the second step is already done as was mentioned above and the first step corresponds to,
for example, the Chapman-Enskog or moment expansion. Then the one-particle distribution function is
replaced by hydrodynamic variables such as the energy density and the fluid velocity.
In the projection operator method, the first step corresponds to the appropriate choices of the projection operators (\ref{moripro}), and the second step corresponds to the coarse graining for the memory function, which we discuss in this section.

\subsection{Violation of Time Reversal Symmetry}

As is commonly believed,
the origin of the violation of the time reversal symmetry is attributed to the existence of two different time scales: one is macroscopic and the other is microscopic.
The time reversal symmetry is violated when the degrees of freedom associated with the latter scale are neglected compared to the former.
However, it is not easy to implement this coarse graining systematically and there are several proposals \cite{anotherapp}.
Here, we discuss the procedure introduced by van Hove in the derivation of the master equation from the Schr\"{o}dinger equation \cite{vanhove}. In his derivation, the time variable $t$ is rescaled as
$\tau$ which is defined by $\tau = g^2 t$ with $g$ being the coupling constant for the microscopic interaction. Then van Hove collected all terms which survive in the asymptotic limit of $t \rightarrow \infty$ with $\tau$ fixed
and succeeded in deriving the quantum master equation. This limit is called the van Hove limit.
Mathematically speaking, the van Hove limit corresponds to the procedure of picking up only secular terms
which dominate the dynamics at the macroscopic time scale (which is realized in small coupling limit in his case).
van Hove's argument indicates that to obtain macroscopic dissipative equation correctly, we have to collect only appropriate terms such as the secular terms.

In the projection operator method, to violate the time reversal symmetry, we have to introduce a coarse graining of time scale in the memory function $\Xi_{ij}(t)$.
As is discussed \cite{mori}, the memory function is given by the time correlation of the noise term,
which is the consequence of the fluctuation-dissipation theorem.
The time scale of the noise is much shorter than that of $\{ A_i \}$
and hence the time scale of $\Xi_{ij}(t)$ can be negligibly small comparing to that of $\{ A_i \}$
\footnote{
If it is not the case, there are two possibilities.
One is that the definition of the projection operator is still incomplete and $\{ A_i \}$
does not span the completely set of the gross variables.
Then we have to generalize the definition of the projection operator.
The other possibility is that there is no coarse-grained macroscopic theory and we have to solve the microscopic dynamics exactly.
Thus the time dependence of the memory function can be the criterion to see whether the definition of the
projection operator is appropriate or not. See also the discussion in Ref. \cite{koidechiral}.
}.
Then the time-convolution integral of Eq. (\ref{morieq}) is approximated as
\begin{equation}
\frac{\partial }{\partial t}\mathbf{A}(t)=i\Delta \ \mathbf{A}%
(t)-\int_{0}^{\infty}d\tau \mathbf{\Xi }(\tau )\mathbf{A}(t) +\mathbf{\xi }%
(t). \label{morieq-1}
\end{equation}
This is the TCL approximation
\footnote{We do not call it the Markov approximation, because the memory effect can exist even after this approximation. For example, the Maxwell-Cattaneo-Vernotte equation which we will discuss later has a memory effect even after the TCL approximation. This is also true for the derivation of transport coefficients of CDRF in the next section.}. See also the discussion in Ref. \cite{hkkr}.
On the other hand, if we carry out the Taylor expansion of ${\bf A}(t-\tau)$ in terms of $\tau$ up to the next-to-leading order \footnote{This is different from the argument developed in Ref. \cite{hkkr}.}, we have
\begin{equation}
\frac{\partial }{\partial t}\mathbf{A}(t)=i\Delta \ \mathbf{A}%
(t)-\int_{0}^{\infty}d\tau \mathbf{\Xi }(\tau ) ( \mathbf{A}(t) -\tau \partial_t \mathbf{A}(t))
+\mathbf{\xi }(t). \label{morieq-2}
\end{equation}%
Note that we still assume that the dominant contribution of the integral comes from $\tau \sim 0$
and the upper limit of the integration is replaced by $\infty$.

At first glance, it might be considered that the latter approach (Eq. (\ref{morieq-2})) is more reliable because the next-order correction in the time-derivative expansion is considered.
However, this is not trivial from the view point of the appropriate violation of the time reversal symmetry.
Here we discuss which approximation is consistent with van Hove's argument \cite{vanhove} by
applying the projection operator method to derive the quantum master equation.
The detailed derivation is shown in Appendix \ref{app1}.
As is shown by Eq. (\ref{ori-naka}),
van Hove's result is reproduced only when the TCL approximation is adopted.
That is, the TCL approximation corresponds to the procedure of collecting all the secular terms in deriving the quantum master equation.
On the other hand, when we consider the next-order correction to the TCL approximation,
we pick up even irrelevant contributions.
This is one of the evidences supporting that the TCL approximation may work better than Eq. (\ref{morieq-2}) in describing macroscopic physics.

\subsection{Exact Relation for Transport Coefficients}

Next, we show that the TCL approximation gives a consistent result with
an exact relation obtained from the microscopic dynamics,
while the next-order correction gives rise to an inconsistency \cite{koide_dif,kadanof,hkkr}.

To show this, let us consider a non-relativistic diffusion process.
Before deriving the (generalized) diffusion equation from the microscopic dynamics, we will point out that
there is an exact relation for the dynamics of a conserved density following Refs. \cite{koide_dif,kadanof,hkkr}.
We consider the complex Schr\"{o}dinger fields, $\psi$ and $\psi^\dagger$,
whose dynamics conserves the (spatial integration of the) number density defined by $n = \psi^{\dagger} \psi$,
as is shown in Ref. \cite{koide_dif}. From the Noether's theorem,
we can obtain the corresponding current operator ${\bf J}$. Then there is the following relation,
\begin{equation}
\langle [{\bf J}({\bf x},t), n({\bf x}',t)] \rangle_{eq} = -\frac{i}{m}\nabla \delta^3({\bf x}-{\bf x}') \langle n ({\bf x}')\rangle_{eq}.
\end{equation}
This leads to the f-sum rule,
\begin{equation}
\frac{1}{{\bf k}^2}\int \frac{d\omega}{\pi} \omega {\rm Im} C^R({\bf k},\omega)
=
-\beta \int d^3 {\bf x}({\bf J}({\bf x},0),{\bf J}({\bf 0},0))
= \frac{1}{m}\langle n({\bf x}=0) \rangle_{eq},
\end{equation}
where $\langle ~~ \rangle_{eq}$ denotes the thermal expectation value.
Here we introduced the retarded Green function $C^R$ through
\begin{equation}
-i \langle [n({\bf x},t),n({\bf x}',t')] \rangle_{eq} \theta(t-t')
= \int \frac{d\omega}{2\pi} \int \frac{d^3 {\bf k}}{(2\pi)^3} C^R ({\bf k},\omega) e^{-i\omega(t-t')}
e^{i{\bf k}({\bf x}-{\bf x}')}.
\end{equation}
By using this relation, the exact form of the Laplace-Fourier transform of the time evolution of the conserved number density $\delta n ({\bf x},t)= n({\bf x},t) - \langle n({\bf x}) \rangle_{eq}$ is given by
\begin{eqnarray}
\frac{\delta n^{LF}({\bf k},z)}{F({\bf k})}
=
\frac{i}{z} C^R ({\bf k},0) - \frac{i}{z^3} {\bf k}^2 \beta \int d^3 {\bf x}({\bf J}({\bf x},0),{\bf J}({\bf 0},0)) + O(1/z^4), \label{diffexa}
\end{eqnarray}
where $F({\bf k})$ is an arbitrary function related to the initial condition.
This is the exact relation which is obtained from quantum mechanics.
Note that the above relation is independent of the form of the interaction term in the Hamiltonian.

As was already emphasized in Refs. \cite{koide_dif,kadanof,hkkr},
if we assume the diffusion equation as the coarse-grained dynamics for $\delta n ({\bf x})$,
we cannot reproduce the exact result (\ref{diffexa}).
In fact, the Laplace-Fourier transform of the diffusion equation is given by
\begin{equation}
\frac{\delta n^{LF}({\bf k},z)}{F({\bf k})}
=
i\frac{C^R({\bf k},0)}{z}
- \frac{C^R({\bf k},0) D {\bf k}^2}{z^2} + \cdots.
\end{equation}
Differently from the exact result (\ref{diffexa}), one observes the following features: 1) there is a $1/z^2$ term which does not disappear in Eq. (\ref{diffexa}) and 2) the coefficient for $1/z^3$ cannot reproduce the exact result, although it is not shown here.
Thus the diffusion equation cannot reproduce the exact behavior obtained from quantum mechanics.

However, if we consider the Maxwell-Cattaneo-Vernotte-type generalized diffusion equation,
\begin{eqnarray}
&& \partial_t \delta n({\bf x},t) + \nabla {\bf J} ({\bf x},t) = 0, \\
&& \partial_t {\bf J}({\bf x},t) = - \frac{D}{\tau_D} \nabla n ({\bf x},t) - \frac{1}{\tau_D} {\bf J} ({\bf x},t),
\end{eqnarray}
we can completely reproduce Eq.(\ref{diffexa}) up to $O(1/z)^3$ as,
\begin{equation}
\frac{\delta n^{LF}({\bf k},z)}{F({\bf k})}
=
i\frac{C^R({\bf k},0)}{z} + i \frac{D{\bf k}^2 C^R({\bf 0},0)}{\tau_D z^3} + \cdots ,
\end{equation}
if the transport coefficients
$D$ and $\tau_D$ satisfy the following relation,
\begin{equation}
\frac{D}{\tau_D} = -\frac{\beta}{C^R ({\bf 0},0)}\int d^3 {\bf x}({\bf J}({\bf x},0),{\bf J}({\bf 0},0))
= \frac{\int d^3 {\bf x}({\bf J}({\bf x},0),{\bf J}({\bf 0},0)) }
{\int d^3 {\bf x}( \delta n ({\bf x},0), \delta n ({\bf 0},0)) } . \label{dif_ex_ratio}
\end{equation}

If we require our approximation to the projection operator method to be consistent with the f-sum rule, the derived $D$ and $\tau_D$ must satisfy Eq. (\ref{dif_ex_ratio}).
In fact, as shown in Ref. \cite{koide_dif}, if define the projection operator with two gross variables, $\delta n({\bf x})$ and ${\bf J}({\bf x})$,
the equations of motion read
\begin{eqnarray}
&& \partial_t \delta n ({\bf k},t) + i{\bf k} \cdot {\bf J} ({\bf k},t) = 0, \nonumber \\
&& \partial_t {\bf J} ({\bf k},t) = - i {\bf k} \Omega^D \delta n ({\bf x},t) - \int^t_0 d\tau \Xi^D({\bf k},\tau) {\bf J}({\bf k},t-\tau) ,
\end{eqnarray}
where
\begin{equation}
\Omega^D = \frac{\int d^3 {\bf x}({\bf J}({\bf x},0),{\bf J}({\bf 0},0)) }
{\int d^3 {\bf x}( \delta n ({\bf x},0), \delta n ({\bf 0},0)) } .
\end{equation}

When we adopt the TCL approximation to the memory function $\Xi^D({\bf k},\tau)$, we find the following relations,
\begin{eqnarray}
\frac{1}{\tau_D} &=& \int^\infty_0 d\tau \Xi^D({\bf 0},\tau), \\
\frac{D}{\tau_D} &=& \Omega^D .
\end{eqnarray}
They coincide exactly with the exact result (\ref{diffexa}).

On the other hand, if we consider the higher order correction as is done in Eq. (\ref{morieq-2}),
the definition of $1/\tau_D$ is modified as is shown in Eq. (65) of Ref. \cite{hkkr},
and one cannot reproduce the exact relation for the ratio $D/\tau_D$ anymore.
This is the second reason of why we do not take into account the next-order correction.

The above argument is developed for specific examples, the quantum master equation and the generalized diffusion equation.
Thus, exactly speaking,
it is not obvious whether the next-order correction still causes inconsistency in the transport coefficients of CDRF.
There may exist examples where the time-reversal symmetry can be violated
by the derivative expansion used to derive Eq. (\ref{morieq-2}).

However, in the case of CDRF, there is another evidence to support the TCL approximation.
The transport coefficients of CDRF can be calculated even from the Boltzmann equation
with the 14 moment approximation \cite{is}.
In Ref. \cite{dhkr},
it was shown that those results are consistent (not the same) with those from
the projection operator when the TCL approximation is employed at vanishing chemical potential.
The fact that the consistent results are obtained from two different approaches is very surprising and strongly suggests to employ the TCL approximation.

The expression of the relaxation time including the next-order correlation is explicitly given by
Eq. (65) in Ref. \cite{hkkr},
and this expression is exactly the same as the result obtained in Ref. \cite{con}.
That is, our projection operator approach can reproduce the result of Ref. \cite{con} by changing the
approximation to the memory function.
However, as was discussed in this section, we cannot obtain reasonable results if we include this
next-order correction to the TCL approximation.
This indicates that the derivative expansion, which is used in Ref. \cite{con},
may not violate the time reversal symmetry appropriately.
In fact, exactly speaking, hydrodynamics derived in this way corresponds
to the relativistic Burnett equation and the corresponding transport
coefficients are not those of CDRF, although it was assumed that they are equal in some literatures.
Note that the Burnett equation has an intrinsic problem called the Bobylev instability and is
essentially different from CDRF \cite{Bobylev,colin}.
Furthermore, the weak coupling limit of the result of Ref. \cite{con} was discussed in Ref.
\cite{moore}, but it is not consistent with the results from the Boltzmann equation
with the 14 moment approximation as is discussed in Ref. \cite{dhkr}.

\section{Transport coefficients at finite chemical potential} \label{fcp}

In this section, we will apply Eq. (\ref{morieq}) and the TCL approximation to
derive microscopic formulas for the shear viscosity, bulk viscosity and the corresponding relaxation times of CDRF.

\subsection{Shear viscosity and corresponding relaxation time}

Following Ref. \cite{knk},
let us consider a fluid flowing in the $x$ direction with
finite velocity-gradient in the $y$ direction. Thus the bulk viscous pressure does not show up.
Then $T^{0x}$ and $T^{yx}$ are chosen as the gross variables, and the projection operator is
given by
\begin{eqnarray}
P O &=& \frac{(O, {T}^{0x}(-k^y))}{( {T}^{0x}(k^y), {T}^{0x}(-k^y))} {T}^{0x}(k^y)
+ \frac{(O, {T}^{yx}(-k^y))}{( {T}^{yx}(k^y), {T}^{yx}(-k^y))} {T}^{yx}(k^y) .
\end{eqnarray}
Here we have implemented the Fourier transformation in space.
Substituting it into Eq. (\ref{morieq}), we obtain
\begin{eqnarray}
\partial_t T^{0x} (k^y,t) &=& -ik_y T^{yx}(k^y,t), \\
\partial_t T^{yx} (k^y,t) &=& -ik_y R^\pi_{k^y} T^{0x}(k^y,t)-\int^t_0 d\tau \Xi^\pi (k^y,\tau)
T^{yx} (k^y, t-\tau).
\end{eqnarray}
One can easily confirm that this equation is still symmetric under the time reversal operation, $t \leftrightarrow -t$,
by using the exact expression for $\Xi^{\pi}(k^y,t)$ \cite{mori,books}.
In order to violate the time reversal symmetry, we employ the coarse graining of time scale by using the TCL approximation,
\begin{eqnarray}
\partial_t T^{0x} (k^y,t) &=& -ik_y T^{yx}(k^y,t), \\
\partial_t T^{yx} (k^y,t) &=& -ik_y R^\pi_{k^y} T^{0x}(k^y,t) - \int^\infty_0 d\tau \Xi^\pi (k^y,\tau)
T^{yx} (k^y, t), \label{sheareq}
\end{eqnarray}
where
\begin{equation}
R^\pi_{k^y} = \frac{({T}^{yx}(k_y), {T}^{yx}(-k_y))}{({T}^{0x}(k_y),{T}^{0x}(-k_y))} .
\end{equation}
As a result, the second term on the right hand side does not contain the time convolution integral any more.
See Refs. \cite{pro,hkkr,koide_dif,mori,pro2} for details.
The Laplace transform of the memory function $\Xi^\pi (k^y,\tau)$ is given by
\begin{equation}
\lim_{s,k^y \rightarrow 0} \Xi^{\pi L} (k^y,s) = \frac{1}{X^{\pi L}(k^y,s)} ,
\end{equation}
with
\begin{equation}
X^{\pi L}(k^y,s)
= \int^\infty_0 dt e^{-st} \frac{(T^{yx}(k^y,t),{T}^{yx}(-k^y))}
{({T}^{yx}(k^y), {T}^{yx}(-k^y))}.
\end{equation}

On the other hand, the phenomenological equation for the shear stress tensor near the rest frame is
\begin{equation}
\tau_\pi \frac{\partial}{\partial t} T^{yx} (k^y)
+ T^{yx} (k^y) = -\eta (ik^y) u^x (k^y), \label{ph_shear}
\end{equation}
where $\eta$ and $\tau_\pi$ are the shear viscosity and the corresponding relaxation time,
respectively.
By comparing
Eq. (\ref{sheareq}) with Eq. (\ref{ph_shear}), we finally obtain the microscopic expressions for
$\eta$ and $\tau_\pi$ as
\begin{subequations}
\begin{eqnarray}
\frac{\eta}{\beta(\varepsilon + P)}
&=&
\lim_{s,k^y \rightarrow 0}\frac{1}{\beta}R^\pi_{k^y} X^{\pi L} (k^y,s)\nonumber\\
&=&
\frac{\displaystyle \lim_{s\rightarrow 0}
\int dt d^3 {\bf x} e^{-st} (T^{yx}({\bf x},t), {T}^{yx}({\bf 0}))}
{\displaystyle \beta \int d^3 {\bf x} ({T}^{0x}({\bf x}),{T}^{0x}({\bf 0})) }, \\
\frac{\tau_\pi}{\beta}
&=&
\lim_{s,k^y \rightarrow 0}\frac{1}{\beta} X^{\pi L} (k^y,s)\nonumber\\
&=&
\frac{\displaystyle \lim_{s\rightarrow 0}
\int dt d^3 {\bf x} e^{-st} (T^{yx}({\bf x},t), {T}^{yx}({\bf 0}))}
{\displaystyle \beta \int d^3 {\bf x} ({T}^{yx}({\bf x}),{T}^{yx}({\bf 0})) }.
\end{eqnarray}
\label{shear_formula}
\end{subequations}
These expressions are formally the same as the previous results obtained at
vanishing chemical potential \cite{knk}.
The difference comes from the fact that, at finite chemical potential,
the time evolution is governed by $H$ but the ensemble average is taken with $K=H - \mu N$.
Thus it should be noted that the numerators of Eq. (\ref{shear_formula}) can be
expressed using the analytic continuation of the Matsubara function only when
$T^{yx}$ and $T^{0x}$ commute with $N$ \cite{fw}.

In general, transport coefficients are related to the imaginary part of certain
retarded Green functions. Thus one cannot obtain finite results for $\eta$ and $\tau_\pi$
unless the effect of interaction is considered.
On the other hand, the dimensionless $\eta$-$\tau_\pi$ ratio defined by $\eta/\tau_\pi(\varepsilon+P)$ is finite even in the non-interacting case, because this ratio is
given by the real part of the Green functions.
Here, for the sake of simplicity, we consider the leading-order estimation, that is,
we use the free gas approximation to calculate this ratio.

Let us first consider a charged scalar boson described by the following Lagrangian,
\begin{eqnarray}
\label{bosonlag}
{\cal L}
&=&\partial_\mu\phi^\dag\partial^\mu\phi-m^2\phi^\dag\phi.
\end{eqnarray}
As is discussed in Refs. \cite{callan,hkkr}, the canonical energy-momentum tensor
for the scalar field theory is not well-defined.
Instead, we use the improved energy-momentum tensor which reads,
\begin{eqnarray}
\label{tmnboson}
T^{\mu\nu}&=&\partial^\mu\phi^\dag\partial^\nu\phi+\partial^\nu\phi^\dag\partial^\mu\phi
\nonumber\\&-&g^{\mu\nu}(\partial_\rho\phi^\dag\partial^\rho\phi-m^2\phi^\dag\phi)
\nonumber\\&-&\frac{1}{3}(\partial^\mu\partial^\nu-g^{\mu\nu}\partial^2)\phi^\dag\phi.
\label{iemtensor}
\end{eqnarray}
Then a straightforward calculation leads to
\begin{equation}
\frac{\eta}{\tau_\pi (\varepsilon + P)}
=
\frac{\int d^3 {\bf x} ({T}^{yx}({\bf x}),{T}^{yx}({\bf 0}))}{\int d^3 {\bf x} ({T}^{0x}({\bf x}),{T}^{0x}({\bf 0}))}
= \frac{P}{\varepsilon + P}, \label{shearboson}
\end{equation}
where the energy density $\varepsilon$ and the pressure $P$ are given by
\begin{eqnarray}
\varepsilon &=& \int \frac{d^3 p}{(2\pi)^3} E_p
\left( \frac{1}{e^{\beta(E_p - \mu)}-1} + \frac{1}{e^{\beta(E_p + \mu)}-1}\right), \label{ener}\\
P &=&  \int \frac{d^3 p}{(2\pi)^3 E_p} \frac{p^2}{3}
\left( \frac{1}{e^{\beta(E_p - \mu)}-1} + \frac{1}{e^{\beta(E_p + \mu)}-1}\right) , \label{press}
\end{eqnarray}
respectively. Here $E_p = \sqrt{p^2 + m^2}$.
In this derivation, the temperature-independent divergent term is neglected.

\begin{figure}[t]
\includegraphics[scale=0.5]{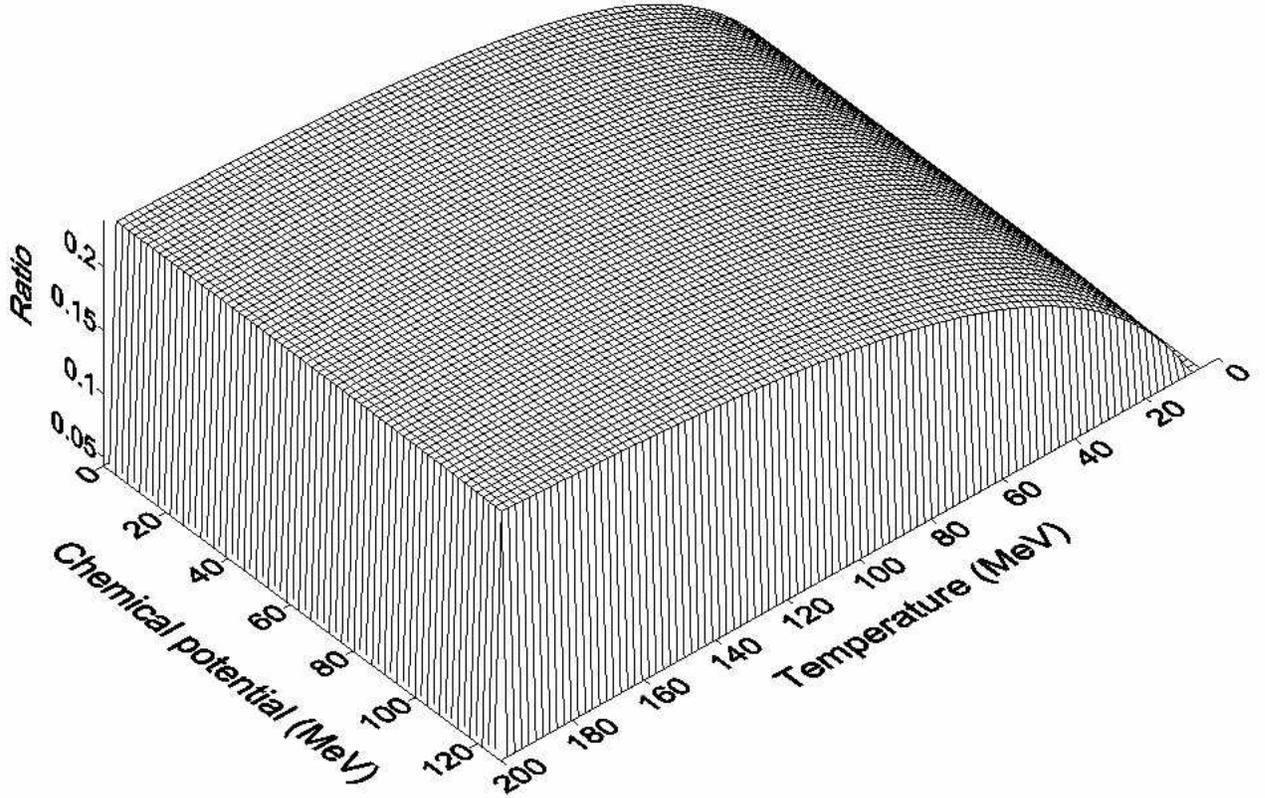}
\caption{The $\eta$-$\tau_\pi$ ratio, $\eta/\tau_\pi(\varepsilon+P)$, as a function of temperature and chemical potential.}
\label{fig:shear1}
\end{figure}

The temperature and chemical potential dependence of the $\eta$-$\tau_\pi$ ratio given by
Eq. (\ref{shearboson}) is shown in Fig. \ref{fig:shear1}. We use the pion mass $m=140$ MeV.
One can see that the ratio monotonically increases as a function of temperature, and
converges to $\frac{P}{\varepsilon+P} = 0.25$ which is the result for the massless bosons.

\begin{figure}[t]
\includegraphics[scale=0.5]{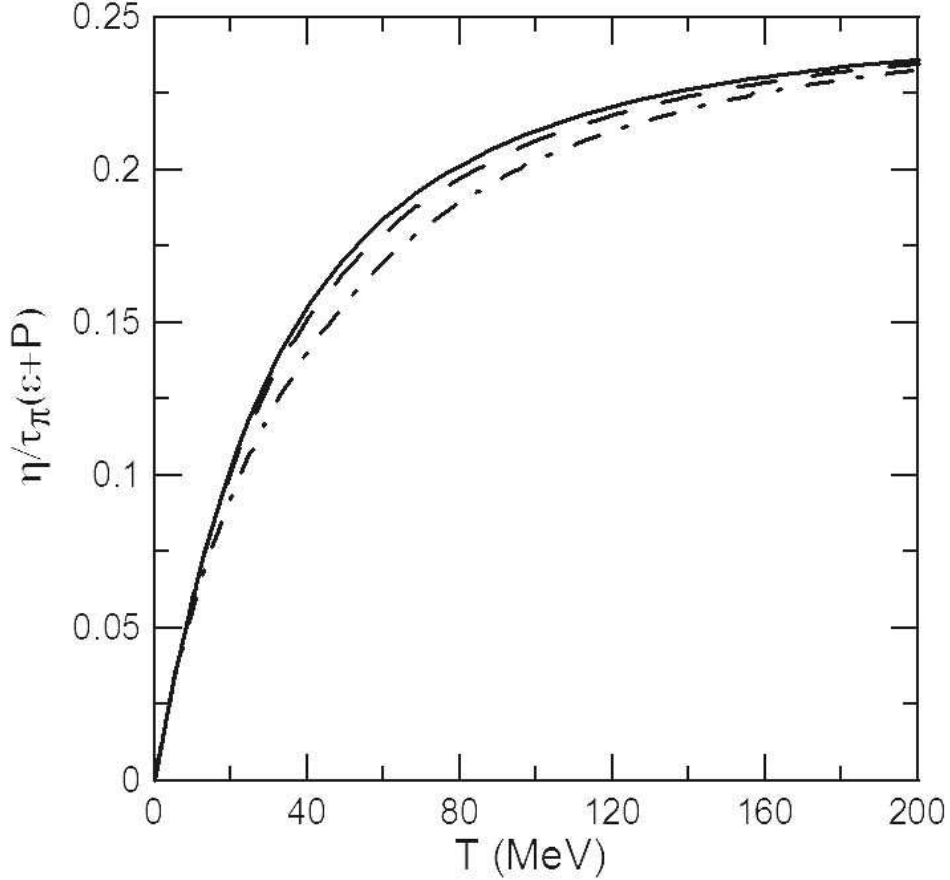}
\caption{The $\eta$-$\tau_\pi$ ratio as a function of temperature at fixed chemical potentials.
The solid, dashed and dot-dashed lines represent the ratio at $\mu=50$, $100$ and $130$ MeV, respectively.}
\label{fig:shear2}
\end{figure}

Compared to the temperature dependence, the chemical potential dependence is not easily recognized. In order to see the chemical potential dependence more clearly,
we plotted the temperature dependence at fixed chemical potentials, $\mu=50$, $100$ and $130$ MeV in Fig.~\ref{fig:shear2}.
One can see that the weak chemical potential dependence still exists and
the $\eta$-$\tau_\pi$ ratio decreases slowly as the chemical potential grows.

As is discussed in Refs. \cite{cau_sta}, this ratio is
directly related to the propagation speed of signals in CDRF.
Thus, for the theory to be relativistically causal, the $\eta$-$\tau_\pi$ ratio should not
be larger than one. This condition is satisfied for any temperature and chemical potential for
our calculation.

These results will be modified by the interaction.
As a matter of fact there are attempts to estimate this ratio including the effect of interaction by using the lattice QCD simulation, see, for example, Ref. \cite{maezawa}.

\subsection{Bulk viscosity and corresponding relaxation time}

Following Ref. \cite{hkkr}, we consider a perturbation in an infinite fluid in thermal
equilibrium having a planar symmetry in the $(y, z)$ plane. All
the quantities associated with the perturbed fluid dynamics
vary spatially only along the $x$ direction. In this case, the fluid
velocity points to the $x$ direction.
Then $T^{0x}$ and $\Pi$ are chosen as the gross variables and the projection operator is given by
\begin{eqnarray}
\label{pobulk}
P O &=& \frac{(O,{T}^{0x}(-k^x))}{({T}^{0x}(k^x),{T}^{0x}(-k^x))} {T}^{0x}(k^x)
\nonumber\\&+& \frac{(O,{\Pi}(-k^x))}{({\Pi}(k^x),{\Pi}(-k^x))} {\Pi}(k^x) .
\end{eqnarray}
Here the operator of the bulk viscous pressure $\Pi$ is defined by the deviation
from the equilibrium pressure.

Differently from the case of the shear stress tensor, the definition of the operator of
the bulk viscous pressure is changed by finite chemical potential.
As is discussed in Ref. \cite{jo,hkkr}, the transport coefficients are expressed by the vanishing momentum limit of commutators. Thus even if we added a term which commutes with the Hamiltonian of our system to the definition of $\Pi$, the final result is not affected except for the arbitrariness related to renormalization.
In fact, from the view point of renormalization, we should add appropriate operators to the original definition of $\Pi$ \cite{jo}.
For the case of the vanishing chemical potential, we used \cite{hkkr}
\begin{equation}
\Pi = \frac{1}{3}\sum_{i=1}^3 (T^{ii} - \langle T^{ii} \rangle_{eq})
- \frac{dP}{d\varepsilon} (T^{00} - \langle T^{00} \rangle_{eq}). \label{def_pi1}
\end{equation}
Note that the last term vanishes in the massless limit and does not violate the conformal property.
The same modification of the bulk viscosity was proposed in Ref. \cite{jo,zubarev}.
As the justification of this modification, we would like to point out that
the bulk viscosity calculated with our microscopic formula is consistent with the result of the Boltzmann equation only when we use Eq. (\ref{def_pi1}) as the operator definition of the bulk viscous pressure.

At finite chemical potential, the operator of the conserved number density $n$ commutes
with the Hamiltonian, and the definition of $\Pi$ can be modified as
\begin{equation}
\Pi = \frac{1}{3} (T^{ii}-\langle T^{ii} \rangle_{eq}) -
\left(\frac{\partial P}{\partial\varepsilon}\right)_n(T^{00} - \langle T^{00} \rangle_{eq})-
\left(\frac{\partial P}{\partial n}\right)_\varepsilon (n-\langle n \rangle_{eq}).
\label{def_pi}
\end{equation}
Note that $\left(\partial P/\partial n\right)_\varepsilon$ vanishes in the massless limit.
See Appendix \ref{app2} for details.
Thus this new definition does not violate the conformal property of the bulk viscous pressure.
The same modification of $\Pi$ was discussed in Refs. \cite{mori,zubarev}.

Substituting Eq. (\ref{pobulk}) into Eq. (\ref{morieq}), we obtain
\begin{eqnarray}
\partial_t T^{0x} (k^x,t) &=& -ik^x \Pi(k^x,t), \\
\partial_t \Pi (k^y,t) &=& -ik^x R^\Pi_{k^x} T^{0x}(k^x,t)-
 \int^t_0 d\tau \Xi^\Pi (k^x,t)
\Pi (k^x, t-\tau), \label{bulkeq}
\end{eqnarray}
where
\begin{equation}
R^\Pi_{k^x} = \frac{({\Pi}(k^x),{\Pi}(-k^x))}{({T}^{0x}(k^x),{T}^{0x}(-k^x))} .
\end{equation}
Similarly to the previous subsection, the TCL approximation is
already employed \cite{hkkr}.
The Laplace transformation of the memory function is given by
\begin{equation}
\lim_{s,k^x \rightarrow 0} \Xi^{\Pi L} (k^x,s) = \frac{1}{X^{\Pi L}(k^x,s)} ,
\end{equation}
where
\begin{equation}
X^{\Pi L}(k^x,s)
= \int^\infty_0 dt e^{-st} \frac{(\Pi(k^x,t),\Pi(-k^x))}
{(\Pi(k^x), \Pi(-k^x))}.
\end{equation}

On the other hand, the phenomenological equation of the bulk viscous pressure
is given by
\begin{equation}
\tau_\Pi \frac{\partial}{\partial t}\Pi(k^x) + \Pi(k^x) =
- \zeta (ik^x) u^x (k^x), \label{ph_bulkeq}
\end{equation}
where $\zeta$ and $\tau_\Pi$ are the bulk viscosity and the corresponding relaxation time,
respectively.

The microscopic expressions for $\zeta$ and $\tau_\Pi$ are obtained by comparing Eq. (\ref{bulkeq})
with Eq. (\ref{ph_bulkeq}), employing the coarse graining of time \cite{hkkr},
\begin{subequations}
\label{bulk_formula}
\begin{eqnarray}
\frac{\zeta}{\beta(\varepsilon + P)}
&=&
\lim_{s,k^x \rightarrow 0}\frac{1}{\beta}R^\Pi_{k^x} X^{\Pi L} (k^x,s)\nonumber\\
&=&
\frac{\displaystyle \lim_{s\rightarrow 0} \int dt d^3 {\bf x} e^{-st} (\Pi({\bf x},t),{\Pi}({\bf 0}))}{\displaystyle \beta \int d^3 {\bf x} ({T}^{0x}({\bf x}),{T}^{0x}({\bf 0})) }, \\
\frac{\tau_\Pi}{\beta}
&=&
\lim_{s,k^x \rightarrow 0}\frac{1}{\beta} X^{\Pi L} (k^x,s)\nonumber\\
&=&
\frac{\displaystyle \lim_{s\rightarrow 0} \int dt d^3 {\bf x} e^{-st} (\Pi({\bf x},t),{\Pi}({\bf 0}))}{\displaystyle \beta \int d^3 {\bf x} ({\Pi}({\bf x}),{\Pi}({\bf 0})) }.
\end{eqnarray}
\end{subequations}

Similarly to the case of the shear viscosity, we calculate the $\zeta$-$\tau_\Pi$ ratio
defined by $\zeta/\tau_\Pi(\varepsilon+P)$ by applying our results to a non-interacting
charged scalar boson.
By using Eqs. (\ref{iemtensor}) and (\ref{def_pi}), we can find $\Pi$.
Then the $\zeta$-$\tau_\Pi$ ratio is calculated as
\begin{eqnarray}
\frac{\zeta}{\tau_\Pi (\varepsilon + P)}
&=&
\frac{\int d^3 {\bf x} ({\Pi}({\bf x}), {\Pi}({\bf 0}))}{\int d^3 {\bf x} (T^{0x}({\bf x}), T^{0x}({\bf 0}))}\nonumber\\
&=& \frac{\left(\frac{1}{3}-c_s^2\right)(\varepsilon+P)-\frac{2}{9}(\varepsilon-3P)}{\varepsilon + P},
\label{bulkboson}
\end{eqnarray}
where the sound velocity $c_s$ is defined by
\begin{eqnarray}
c_s^2=\left(\frac{\partial P}{\partial\varepsilon}\right)_{\frac{s}{n}}=\left(\frac{\partial P}{\partial\varepsilon}\right)_{n}+\frac{n}{\varepsilon+P}\left(\frac{\partial P}{\partial n}\right)_{\varepsilon},
\end{eqnarray}
where $\varepsilon$ and $P$ are defined in Refs. (\ref{ener}) and ({\ref{press}}),
respectively, and
\begin{equation}
n = \int \frac{d^3 p}{(2\pi)^3} \left( \frac{1}{e^{\beta(E_p - \mu)}-1}-\frac{1}{e^{\beta(E_p + \mu)}-1} \right).
\end{equation}
In this derivation, we neglected the temperature independent divergent term.
One can easily check that the $\zeta$-$\tau_\Pi$ ratio disappears in the massless limit.

Differently from the case of the vanishing chemical potential,
this ratio at the finite chemical potential has not yet been calculated
in the frame work of coupled Boltzmann equation with particles and anti-particles.
Thus we cannot discuss the validity of this result from the view point of consistency with
the kinetic theory.
However, we find that the functional form of this ratio is
completely equivalent to the
result for the vanishing chemical potential shown in Ref. \cite{hkkr}.
This is not observed if we ignore the last term of Eq. (\ref{def_pi}) which is newly added.
Furthermore, the functional form of the $\eta$-$\tau_\pi$ ratio is not changed by
the introduction of the chemical potential as was shown in Eq. (\ref{shearboson}).
Because of these facts, we believe that our result is reasonable and
the bulk viscous pressure should be defined by Eq. (\ref{def_pi}) at finite chemical potential.

\begin{figure}[t]
\includegraphics[scale=0.5]{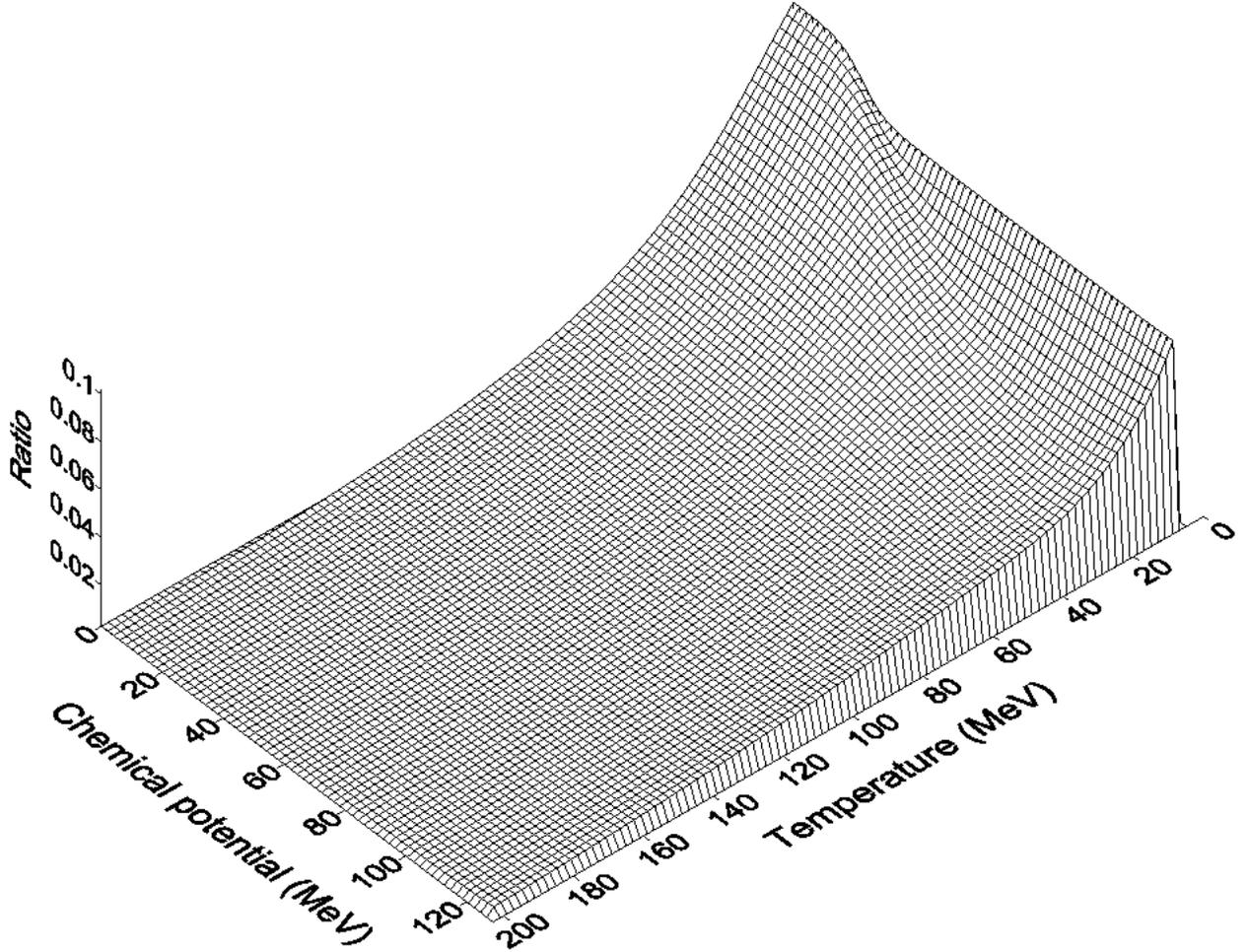}
\caption{The $\zeta$-$\tau_\Pi$ ratio, $\zeta/\tau_\Pi(\varepsilon+P)$, as a function of temperature and chemical potential.}
\label{fig:bulk1}
\end{figure}

The temperature and chemical potential dependence of the $\zeta$-$\tau_\Pi$ ratio given by
Eq. (\ref{bulkboson}) is shown in Fig. \ref{fig:bulk1}. We used the pion mass $m=140$ MeV.
One can see that the $\zeta$-$\tau_\Pi$ ratio monotonically decreases as a function of temperature, and
finally vanishes.
This is because, at very high temperature,
the existence of mass is negligible and then the
system effectively restores the conformal symmetry.

\begin{figure}[t]
\includegraphics[scale=0.5]{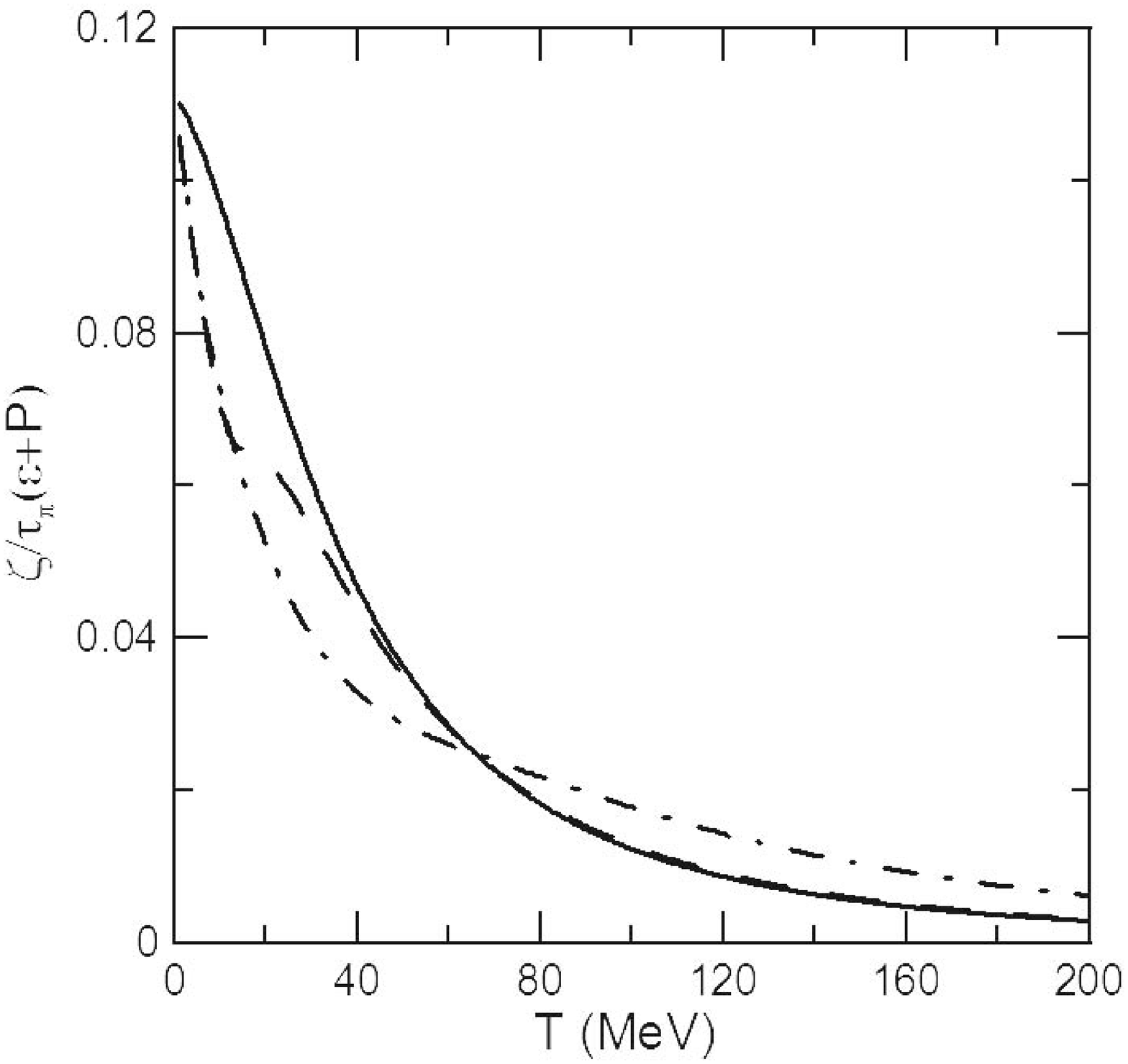}
\caption{The $\zeta$-$\tau_\Pi$ ratio as a function of temperature at fixed chemical potentials.
The solid, dashed and dot-dashed lines represent the ratio at $\mu=0$, $50$ and $130$ MeV, respectively.}
\label{fig:bulk2}
\end{figure}

In Fig. \ref{fig:bulk2},
the temperature dependence of the $\zeta$-$\tau_\Pi$ ratio is plotted
at fixed chemical potentials, $\mu =0$, $50$ and $130$ MeV.
At $\mu=0$, the ratio exhibits maximum at the vanishing chemical potential.
However, it decreases more quickly for smaller chemical potentials.
Then at higher temperature, this ratio becomes larger as the chemical potential increases.

Similarly to the case of the shear viscosity, this ratio is
directly related to the propagation speed of signals in CDRF.
Thus, for the theory to be relativistically causal, the ratio should not be larger than
one. This condition is satisfied for any temperature and chemical potential for our results.

\subsection{fermion}

So far, we have discussed the non-interacting charged scalar bosons.
Parallelly, we can apply the same calculations to non-interacting fermions.

Then we obtain
\begin{subequations}
\begin{eqnarray}
\label{taupif}
\frac{\eta}{\tau_\pi(\varepsilon + P)}
&=&0, \\
\frac{\zeta}{\tau_\Pi (\varepsilon + P)}
&=& \frac{\left(\frac{1}{3}-c_s^2\right)(\varepsilon+P)-\frac{1}{3}(\varepsilon-3P)}{\varepsilon + P}.
\end{eqnarray}
\end{subequations}
There results are the same as those obtained for the vanishing chemical potential case \cite{dhkr} . The reason
for the vanishing of $\eta/{\tau_\pi(\varepsilon + P)}$ for fermions is because that the contribution to
this quantity from the pair annihilation and
creation (PAC) processes cancels the contribution from non-PAC processes. More discussion
can be found in Ref. \cite{dhkr}.

As is emphasized in Ref. \cite{dhkr},
these calculations are performed at leading order and will be modified by the effect of interactions.
For example, the exact expressions for $\tau_\pi /\beta$ and $\tau_\Pi/\beta$
are given by the ratio of the real and imaginary parts
of the retarded Green's function of $T^{yx}$ and $\Pi$, respectively \cite{hkkr}.
To leading order,
the real part is given by the result of the free-gas approximation,
while the imaginary part is not.
That is, the leading-order calculation violates this exact relation and may lead to
inconsistent results.
The vanishing $\eta/(\tau_\pi (\varepsilon + P))$ for fermions
could also be rendered finite by a more complete calculation including interactions.

In addition, usually fermions interact by exchanging bosons.
In such a mixed system of bosons and fermions, $\eta/(\tau_\pi (\varepsilon + P))$ is given by the contributions from both fermions and bosons and takes a finite value
even if the contribution from fermions vanishes.

\section{Concluding remarks} \label{sum}

In this paper, the microscopic formulas for the shear viscosity  $\eta$,
the bulk viscosity $\zeta$, and corresponding relaxation times $\tau_\pi$ and $\tau_\Pi$
of causal dissipative relativistic fluid-dynamics are obtained at finite temperature and chemical potential.

Before obtaining these formulas, we first discussed the theoretical and mathematical meanings of the TCL approximation, which is used to violate the time reversal symmetry possessed by microscopic dynamics.
 To examine the validity of the TCL approximation, we applied the projection operator method to the derivation of the quantum master equation and to the generalized diffusion equations.
By using the TCL approximation, we could collect only the secular terms appropriately and succeeded in
deriving the quantum master equation obtained by van Hove.
This means that the TCL approximation can pick up appropriate secular terms which should be kept for deriving dissipative equations.
When the TCL approximation is applied to derive a coarse-grained equation of the non-relativistic diffusion process,
the diffusion coefficient and the corresponding relaxation time satisfy the exact relation,
the f-sum rule, which is obtained from quantum mechanics.
On the other hand, we cannot reproduce these results,
if we consider the next-order time-derivative correction to the TCL approximation.

Moreover, as is discussed in Ref. \cite{dhkr}, the shear and bulk viscosities and corresponding relaxation times obtained by using the projection operator method
can be consistent with the results from the Boltzmann equation only when the TCL approximation is applied.
Thus we conclude that, to violate the time reversal symmetry appropriately, we should apply the TCL approximation and should not consider the next-order correction.

However it is worth mentioning that, if we take into account this next-order correction to the TCL approximation,
we obtain the result given by Eq. (65) in Ref. \cite{hkkr}, which is the same formula obtained in Ref. \cite{con}.
The quantitative difference of our result
and theirs are shown in Ref. \cite{hkkr,dhkr} for the vanishing chemical potential.

Next, we derived the formulas of the transport coefficients of
the causal dissipative relativistic fluid dynamics
at finite temperature and chemical potential by using the TCL approximated projection operator method.
The formulas for the shear and bulk viscosities are given by Eqs. (\ref{shear_formula}) and
(\ref{bulk_formula}), respectively.

In the calculation of the bulk viscosity, the operator expression
of the bulk viscous pressure should be chosen appropriately.
In this work, we applied Eq. (\ref{def_pi}), the last term of which does not exist
in the calculation at the vanishing chemical potential \cite{hkkr}.

The transport coefficients $\eta$, $\zeta$, $\tau_\pi$ and $\tau_\Pi$ are
not finite in the free-gas approximation because they are proportional to the
imaginary part of the Green functions.
However, the ratios $\eta/\tau_\pi$ and $\zeta/\tau_\Pi$
are still finite because those are calculated from the real parts of the Green functions.
Then we obtain
\begin{eqnarray}
\label{sh}
\frac{\eta}{\tau_\pi}&=&(3-\alpha)P,\\
\label{bu}
\frac{\zeta}{\tau_\Pi}&=&\left(\frac{1}{3}-c_s^2\right)(\varepsilon+P)-\frac{\alpha}{9}(\varepsilon-3P),
\end{eqnarray}
where $\alpha=2$ for the charged scalar boson and $\alpha=3$ for fermion.
The functional forms of these ratios are
completely the same as the
results for the vanishing chemical potential shown in Ref. \cite{knk,hkkr}.
This equivalence is not observed if we ignore the last term of Eq. (\ref{def_pi}) which is newly added.
It should be noted that these expressions are confirmed only for the leading order calculations and
it has not yet known how these relations are modified under the effects of interaction.

This may suggest the consistency of our calculations.
However, to confirm it more precisely, we should compare our results with those
from the Boltzmann equation including particles and anti-particles.
In the case of the vanishing chemical potential,
the same ratios were calculated from the simple Boltzmann equation and confirmed that
these are consistent when quantum correlations which are not included
in the Boltzmann equation are neglected \cite{dhkr}.
The consistency check for the case of finite chemical potential is left for future task.

Besides the viscosities and corresponding relaxation times, CDRF {\it can} contain other transport coefficients.
For example, the heat conductivity and the corresponding relaxation time
are other important transport coefficients \footnote{As a matter of fact, the moment expansion
of the Boltzmann equation predicts a lot of non-linear terms. However,
we have to carefully pick up only terms which are consistent from the viewpoint of the
order of the Knudsen number \cite{woods}. }.
These are under investigation.

\hspace{1cm}

{\bf Acknowledgments:}
We thank the support from the Helmholtz International Center for FAIR within the
framework of the LOEWE program (Landesoffensive zur Entwicklung
Wissenschaftlich- \"Okonomischer Exzellenz) launched by the State of Hesse.
T.K. is also financially supported by CNPq.

\appendix

\section{The van Hove limit and the TCL approximation} \label{app1}

In order to see the relation between the van Hove limit and the TCL approximation,
we reformulate the result of van Hove \cite{vanhove} by using the projection operator method,
following the argument of Nakajima \cite{nakajima} and Zwanzig \cite{zwanzig1960}.

We consider a quantum system whose Hamiltonian is given by
\begin{equation}
H = H_0 + g V, \label{ham}
\end{equation}
where $H_0$ is the non-perturbed Hamiltonian and the
Fock space is defined by the eigenstates of $H_0$,
\begin{eqnarray}
H_0 | E \alpha \rangle = E_\alpha | E \alpha \rangle, ~~~~\langle E \alpha | E'\alpha'\rangle =
\delta (E_\alpha-E'_{\alpha'})\delta (\alpha - \alpha'),
\end{eqnarray}
where $\alpha$ is a quantum number except for energy.
Without loss of generality, we assume that the interaction $g V$
contains only off-diagonal parts in this representation \cite{vanhove}.

The dynamics of the density matrix is given by
\begin{equation}
\partial_t \rho (t) = -iL \rho (t).
\end{equation}
This equation is re-expressed as the following form by introducing the general projection
operator $P$ satisfying $P^2 = P$,
\begin{equation}
\label{naka-app}
\partial_tP\rho(t) = -iPL P\rho (t) - iPL e^{-QiL t }Q \rho(0)
+  iPL \int^t_0 ds e^{-QiL s} Q iL P \rho (t-s) ,
\end{equation}
where $Q=1-P$. This is easily derived by using the same method used in deriving
Eq. (\ref{morieq}) or using the following operator identity,
\begin{equation}
e^{-iLt}=Pe^{-iLt}+e^{-iQLt}Q-i\int^t_0 ds e^{-QiL s} Q L P e^{-iL(t-s)}.
\end{equation}

To reproduce the master equation obtained by van Hove, following Ref. \cite{nakajima},
we specify $P$ as a projection operator
which extracts the diagonal parts of an arbitrary operator $A$,
\begin{equation}
\langle E_\alpha,\alpha|PA|E'_{\alpha'},\alpha'\rangle=
\langle E_\alpha,\alpha|A|E'_{\alpha'},\alpha'\rangle\delta(E_\alpha-E'_{\alpha'})\delta(\alpha-\alpha').
\end{equation}
Note that the projection operator defined above is a superoperator.
Following Ref. \cite{vanhove}, we choose the initial condition so that $\rho(0)$ does not
contain off-diagonal components, {\it i.e.}, $Q\rho(0)=0$.
Then Eq. (\ref{naka-app}) becomes
\begin{equation}
\label{naka-app2}
\partial_t\rho_D(t) = -iPL \rho_D (t)
+  iPL \int^t_0 ds e^{-QiL s} Q iL \rho_D (t-s) ,
\end{equation}
where $\rho_D(t)=P\rho(t)$ is the diagonal part of $\rho(t)$.

Corresponding to the Hamiltonian (\ref{ham}), the operator $L$ is separated as
\begin{equation}
L=L_0+L_I,
\end{equation}
where $L_0$ and $L_I$ are the Liouville operators with $H_0$ and $gV$, respectively.
One can easily check that $PL P A=0$ and $PL_0 A =0$ because of the definition of $P$.
Then, from Eq. (\ref{naka-app2}), the equation for $\rho_D (t)$ is re-expressed as
\begin{equation}
\label{eqrhod}
\partial_t \rho_D(t) = P iL_I \int^t_0 ds e^{-QiL s} Q iL_I \rho_D (t-s)
\approx P iL_I \int^t_0 ds e^{-QiL_0 s} Q iL_I \rho_D (t-s).
\end{equation}
Note that, to compare the result of van Hove, it is enough to expand $e^{-QiL s}$ in terms
of the interaction strength $g$ and keep only the lowest order term
(more exactly, these higher order terms disappear in the van Hove limit).
About this expansion, see, for example, Ref. \cite{pro2}.

Let us introduce a probability
$P_E(\alpha,t)\equiv \langle E \alpha | \rho (t)| E \alpha \rangle$
for a particle to be found at state $\alpha$ at time $t$.
Then the equation for this probability is
\begin{eqnarray}
\partial_t P_E(\alpha,t)
&=& -\int^t_0 ds \int d E'_{\alpha'} d{\alpha'} 2g^2
|\langle E \alpha | V |E' \alpha'\rangle|^2 \cos(E'_{\alpha'} - E_{\alpha})s
\left\{
 P_E(\alpha,t-s) - P_{E'}(\alpha',t-s)
\right\} \nonumber \\
&\approx& -\int^\infty_0 ds \int d E'_{\alpha'} d{\alpha'} 2g^2
|\langle E \alpha | V |E' \alpha'\rangle|^2 \cos(E'_{\alpha'} - E_{\alpha})s
\left\{
 P_E(\alpha,t) - P_{E'}(\alpha',t)
\right\} \nonumber \\
&=& \int d E'_{\alpha'} d{\alpha'} 2g^2
|\langle E \alpha | V |E' \alpha'\rangle |^2
\left(
\lim_{t'\rightarrow \infty}\frac{\sin(E'_{\alpha'} - E_\alpha)t'}{E'_{\alpha'}-E_\alpha}
\right)
\left\{
 P_E(\alpha,t) - P_{E'}(\alpha',t)
\right\} \nonumber \\
&=&  \int dE'_{\alpha'} d{\alpha'}w(\alpha;\alpha') \left\{
P_E(\alpha',t) - P_E(\alpha,t)
\right\}, \label{ori-naka}
\end{eqnarray}
where
\begin{equation}
w(\alpha;\alpha') = 2 \pi g^2 |\langle E\alpha | V | E\alpha'\rangle |^2
\delta (E_\alpha - E'_{\alpha'}).
\end{equation}
From the first to the second line on the r.h.s., we used the TCL approximation by replacing the
upper limit of the integration from $t$ to $\infty$, and the index of $P_E$ from $t-s$ to $t$.
The last result is completely equivalent to the master equation obtained by van Hove
by using a different method \cite{vanhove}.
As is well known, the transition matrix $w$ coincides with Fermi's golden rule.

As was mentioned in the text, van Hove obtained this result by collecting all the secular terms
under the van Hove limit. In the projection operator method,
the same result is reproduced using the TCL approximation.
That is, the TCL approximation plays the same role as the van Hove limit
and collects only relevant terms in the coarse graining of time scale.
However, if we consider the next-order correction to the TCL approximation in Eq. (\ref{ori-naka}), we cannot reproduce van Hove's result.

\section{thermodynamic relations} \label{app2}

We will show that $(\partial P/\partial n)_{\varepsilon}$ vanishes in the massless limit.

Note that
\begin{eqnarray}
dP
&=& \left( \frac{\partial P}{\partial \varepsilon} \right)_n d\varepsilon
+ \left( \frac{\partial P}{\partial n} \right)_\varepsilon dn \nonumber \\
&=& sdT + nd\mu.
\end{eqnarray}
From this relation, we obtain
\begin{eqnarray}
\left(
\begin{array}{c}
\left( \frac{\partial P}{\partial \varepsilon} \right)_n \\
\left( \frac{\partial P}{\partial n} \right)_\varepsilon
\end{array}
\right)
=
\frac{1}{Det}
\left(
\begin{array}{cc}
\left( \frac{\partial n}{\partial \mu} \right)_T
& -\left( \frac{\partial n}{\partial T} \right)_\mu \\
-\mu \left( \frac{\partial n}{\partial \mu} \right)_T
- T \left( \frac{\partial s}{\partial \mu} \right)_T
&\mu \left( \frac{\partial n}{\partial T} \right)_\mu
+ T \left( \frac{\partial s}{\partial T} \right)_\mu
\end{array}
\right)
\left(
\begin{array}{c}
s \\
n
\end{array}
\right),
\end{eqnarray}
where
\begin{eqnarray}
Det = \left( \frac{\partial n}{\partial \mu} \right)_T
\left[
\mu \left( \frac{\partial n}{\partial T} \right)_\mu
+ T \left( \frac{\partial s}{\partial T} \right)_\mu
\right]
- \left( \frac{\partial n}{\partial T} \right)_\mu
\left[
\mu \left( \frac{\partial n}{\partial \mu} \right)_T
+ T \left( \frac{\partial s}{\partial \mu} \right)_T
\right].
\end{eqnarray}
Thus $(\partial P/\partial n)_{\varepsilon}$ is expressed as
\begin{eqnarray}
\left(\frac{\partial P}{\partial n}\right)_{\varepsilon}
&=& \left[-s \left(\frac{\partial (\varepsilon+P)}{\partial\mu} \right)_T
+n\left(\frac{\partial (\varepsilon+P)}{\partial T} \right)_\mu \right]\frac{1}{Det}\nonumber\\
&=& \left[ -s\left( \frac{\partial\varepsilon }{\partial\mu} \right)_{T}
+n \left(\frac{\partial\varepsilon}{\partial T} \right)_\mu \right]\frac{1}{Det}.
\end{eqnarray}
In the conformal limit where $\varepsilon=3P$,
$\left( \partial\varepsilon /\partial \mu \right)_{T} = 3 n $
and $\left( \partial\varepsilon/\partial T \right)_\mu = 3s $.
Thus $(\partial P/\partial n)_{\varepsilon}$ vanishes.

\end{document}